\documentclass[prb,twocolumn,amsmath,amssymb,showpacs]{revtex4}
\usepackage{graphicx}

\def\beq{\begin{equation}}
\def\eeq{\end{equation}}
\def\beqa{\begin{eqnarray}}
\def\eeqa{\end{eqnarray}}

\begin{document}


\title{
One-electron self-energies and spectral functions for the
$t-J$ model in the
large-$N$ limit.
}

\author{
M. Bejas, A. Greco and A. Foussats
}

\affiliation{
Facultad de Ciencias Exactas, Ingenier\'{\i}a y Agrimensura
and Instituto de F\'{\i}sica Rosario
(UNR-CONICET).
Av. Pellegrini 250-2000 Rosario-Argentina.\\
}

\date{\today}

\begin{abstract}
Using a recently  developed perturbative approach,  which
considers Hubbard operators as fundamental excitations, we have
performed electronic self-energy and spectral function
calculations for the $t-J$ model on the square lattice. 
We have found that the
spectral functions along the Fermi surface are isotropic, even
close to the critical doping where the $d$-density wave phase
takes place. Fermi liquid behavior with scattering rate $\sim
\omega^2$  and a finite quasiparticle weight $Z$ was obtained. $Z$
decreases with decreasing doping taking low values for low doping.
Results
are compared with other ones, analytical and numerical like
slave-boson and Lanczos diagonalization finding agreement. 
We discuss our results in the light of recent $ARPES$ experiments
in cuprates.
\end{abstract}


\maketitle

\section{Introduction}

Since the discovery of high Tc superconductivity \cite{Muller} a
large part of the solid state community accepted that the $t-J$
model is fundamental for understanding the physics of cuprates
\cite{Anderson}. However, in spite of the great deal of work done, important
questions about this model are still open. In particular, the one-electron
spectral function is one of the most relevant among them. The recent 
improvement of $ARPES$ experiments have allowed to take access to more refined 
information about 
one-electron spectral functions and self-energy effects
\cite{Reviwes}, both considered relevant for understanding the
physics of cuprates.

The main problem for calculating spectral properties in the
framework of $t-J$ model is the treatment, in a controllable way,
of the non-double occupancy constraint. There are several
analytical and numerical approaches used to treat the constrained
algebra of the $t-J$ model. We will mention some of them. From
the analytical side can be mentioned: a) Self consistent Born
approximation \cite{Horsch} which is appropriate for the one-hole
problem. b) Slave fermion (SF) \cite{Izyumov}, even if it is accepted 
that the method is reasonable for low doping, there are
no many calculations of spectral functions which require the
evaluation of fluctuations above the mean field level. c) Slave
boson (SB) \cite{Izyumov}, unlike the SF,  seems to be
appropriate for describing the metallic regime. However, like SF,
the treatment of fluctuations above the mean field  is not trivial
(we further discuss this point below). From the numerical side can
be mentioned: a) Quantum Monte-Carlo (QMC), which is suitable for
calculating spectral functions for one-hole case \cite{Muramatsu}
while for finite doping the sign problem makes  the calculation
uncontrolled. b) Lanczos diagonalization \cite{Dagotto} and its finite
temperature version \cite{Prelov} which are limited to
finite clusters. As a consequence, there is no a single method
covering all situations, therefore it is important to complement
analytical with numerical methods and viceversa.

In Ref. \onlinecite{Foussats} we have developed, for $J=0$, a
perturbative large-$N$ approach for the $t-J$ model based on the
path integral representation for Hubbard-operators (or
$X$-operators) which, in what follows, will be called PIH method.
This method deals with X-operators as fundamental objects and is
not based on any decoupling scheme; thus, there are no
complications as gauge fixing and bose condensation like in the SB
approach \cite{Arrigoni}. Recently \cite{Foussats1}, the PIH
approach was extended to the case of finite $J$.   The obtained
phase diagram and charge-correlations were compared with other
calculations based on SB \cite{Wang} and   Bayn Kadanoff
functional theory (BKF) \cite{Zeyher} finding good agreement.

The aim of the present  paper is to present
one-particle spectral
function and self-energy calculations using
the PIH approach.
We show that the method is useful for explicit
calculation of spectral properties, enabling to sum-up,
systematically, fluctuations above mean field solution given reliable results.

The paper is organized as follows.
In section II, after a brief summary of the PIH method,
we show the analytical expressions for the self-energy and
the spectral function.
In section III, the results are compared with available
SB ones.
In section IV, the results are compared with Lanczos diagonalization ones 
for different ${{\bf{k}}}$'s on the Brillouin zone (BZ)
and different doping levels. 
In section V
we present a detailed analysis for self-energies
and spectral functions for several doping levels and $J$.
Conclusions and discussions are given in section VI.

\section{Brief summary of the formalism. Self-energy and spectral
function calculation}

PIH approach
was developed extensively
in previous
papers\cite{Foussats,Merino,Foussats1,Merino1} and in this section we list the main
useful steps for explicit calculations of the self-energy and spectral functions.

We associate with the $N$-component fermion field $f_{p}$
the propagator,
connecting two generic components $p$ and $p'$,

\begin{eqnarray}\label{G0}
G^{(0)}_{pp'}({\bf k}, \omega_{n}) = - \;
\frac{\delta_{pp'}}{i\omega_{n} - E_{k}}
\end{eqnarray}

\noindent which is $O(1)$. The original spin index $\sigma=\pm$
was extended to the index $p$ running from $1$ to $N$.

The fermion variable $f_{ip}$ is proportional to the fermionic
$X$-operator $X^{0p}_i$, $f_{ip}=(1/\sqrt{Nr_0})X^{0p}_i$, and can
not be associated with the spinons as in SB. In Eq. (\ref{G0}),
$E_{k}$ ($E_{k} = -2(tr_{0}+\Delta) (cosk_x+cosk_y) - \mu$) is the
electronic dispersion in  leading order, where $t$ is the hopping
between nearest-neighbors sites on the square lattice and $\mu$
the chemical potential. The mean field values $r_0$ and $\Delta$
must be determined minimizing the leading order theory. From the
completeness condition ($\sum_p X^{pp}_i+X^{00}_i=N/2$), $r_0$ is
equal to $\delta/2$ where $\delta$ is the hole doping away from
half-filling. The expression for $\Delta$ is

\beq {\label {Delta}}
\Delta = \frac{J}{2 N_s} \sum_k cos(k_x) n_F(E_k)
\eeq

\noindent where $n_F$ is the Fermi function, $J$ the exchange
interaction between nearest-neighbors and $N_s$ is the number of
sites in the BZ. For a given doping $\delta$; $\mu$ and $\Delta$
must be determined self-consistently from
$(1-\delta)=\frac{2}{N_s} \sum_{k} n_F(E_k)$ and Eq. (\ref{Delta}).

\begin{figure}
\vspace{1cm}
\begin{center}
\setlength{\unitlength}{1cm}
\includegraphics[width=8cm,angle=0]{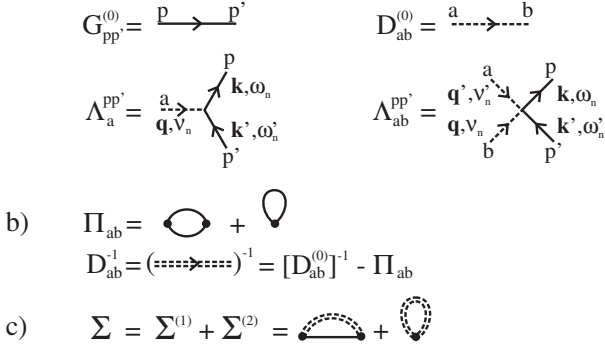}
\end{center}
\caption{a) Summary of the Feynman rules. Solid line represents
the propagator $G^{(0)}$ (Eq.(1)) for the correlated fermion
$f_p$. Dashed line represents the $6 \times 6$ boson propagator
$D^{(0)}$ (Eq. (3)) for the $6$-component field $\delta X^a$. Note
that the component $(1,1)$ of this propagator is directly
associated with the $X^{00}$ charge operator. $\Lambda^{pp'}_a$
(Eq. (4)) and $\Lambda^{pp'}_{ab}$ represent the interaction
between two fermions $f_p$ and one and two bosons $\delta X^a$
respectively. b) $\Pi_{ab}$ contributions to the irreducible boson
self-energy. c) Contributions to the electron self-energy
$\Sigma({\bf k},\omega)$ through $O(1/N)$.}  \label{FD}
\end{figure}

We associate with the six component boson field $\delta X^{a}
= (\delta R\;,\;\delta{\lambda},\; r^{x},\;r^{y},\; A^{x},\;
A^{y})$, the inverse of the propagator,
connecting two generic components $a$ and $b$,

\begin{widetext}
\begin{eqnarray}\label{D0}
D^{-1}_{(0) ab}({\bf q},\nu_{n})= N \left(
 \begin{array}{cccccc}
-2Jr_{0}^{2}(\cos(q_{x})+\cos(q_{y}))
& r_{0} & 0 & 0 & 0 & 0 \\
   r_{0} & 0 & 0 & 0 & 0 & 0 \\
   0 & 0 & \frac{4}{J}\Delta^{2} & 0 & 0 & 0 \\
   0 & 0 & 0 & \frac{4}{J}\Delta^{2} & 0 & 0 \\
   0 & 0 & 0 & 0 & \frac{4}{J}\Delta^{2} & 0 \\
   0 & 0 & 0 & 0 & 0 & \frac{4}{J}\Delta^{2} \
 \end{array}
\right).
\end{eqnarray}
\end{widetext}

\noindent

The bare boson propagator $D^{(0)}_{ab}$ is $O(1/N)$.
The first component $\delta R$ of the $\delta X^a$ field
is related to charge fluctuations by $X_i^{00}=Nr_0(1+\delta R_i)$,
where $X^{00}$ is the Hubbard operator associated with the number of
holes.
$\delta \lambda$ is the fluctuation of the
Lagrangian multiplier $\lambda_i$ associated
with the completeness condition.
$r_i^{\eta}$ and $A_i^{\eta}$ correspond, respectively,
to the amplitude and the
phase fluctuations of the bond variable $\Delta_i^{\eta} = \Delta
(1+r_i^{\eta}+iA_i^{\eta})$ where $\eta = x,y$.

\begin{widetext}

The three-leg vertex,

\begin{eqnarray}\label{gamm3}
\Lambda^{pp'}_{a}& =&  (-1) \left(\frac{i}{2}(\omega_n +
{\omega'}_n) + \mu + 2\Delta \sum_{\eta}
cos(k_\eta-\frac{q_\eta}{2})\; cos\frac{q_\eta}{2};\;1;
- 2\;\Delta\; cos(k_x-\frac{q_x}{2}) \; ; \right.\nonumber \\
&-& \left. 2\;\Delta\;
cos(k_y-\frac{q_y}{2}) ;\right.
\left. 2\; \Delta\; sin(k_x-\frac{q_x}{2}) \;;2 \Delta\;
sin(k_y-\frac{q_y}{2}) \right) \delta^{pp'},
\end{eqnarray}

\noindent
represents the interaction between two fermions and one boson.\\

The four-leg vertex, $\Lambda^{pp'}_{ab}$, represents the interaction between
two fermions and two bosons. The only elements different from zero are:

\begin{eqnarray}
\Lambda^{pp'}_{\delta R \delta R}& =&  \left(\frac{i}{2} (\omega_n + {\omega'}_n)
+ \mu
+ \Delta\; \sum_{\eta}
cos(k_\eta-\frac{q_\eta+q'_\eta}{2})\;
[ cos\frac{q_\eta}{2} \; cos\frac{q'_\eta}{2}\;
+\;cos\frac{q_\eta+q'_\eta}{2}] \right)
\delta^{pp'},
\end{eqnarray}
\end{widetext}

\begin{eqnarray}
\Lambda^{pp'}_{\delta R \delta\lambda}=\frac{1}{2}
\;\delta^{pp'},
\end{eqnarray}

\begin{eqnarray}
\Lambda^{pp'}_{\delta R \; r^{\eta}}= -\Delta \;
cos(k_\eta-\frac{q_\eta+q'_\eta}{2})\; cos\frac{q'_\eta}{2}
\;\delta^{pp'},
\end{eqnarray}

\noindent and

\begin{eqnarray}
\Lambda^{pp'}_{\delta R \; A ^{\eta}}= \Delta \;
sin( k_\eta-\frac{q_\eta+q'_\eta}{2})\; cos\frac{q'_\eta}{2}
\;\delta^{pp'}.
\end{eqnarray}

\noindent Each vertex conserves the momentum and energy and they are $O(1)$.
In  each diagram there is a minus sign for each fermion loop and a
topological factor. A brief summary of the Feynman rules is given
in Fig. \ref{FD}(a). As usual in a large-$N$ approach, any  physical
quantity can be calculated at a given order just by counting the
powers in $1/N$ of vertices and propagators involved in the
corresponding diagrams.

The bare boson propagator $D^{(0)}_{ab}$ is renormalized in
$O(1/N)$. From the Dyson equation, $(D_{ab})^{-1} =
(D^{(0)}_{ab})^{-1} - \Pi_{ab}$, the dressed  components $D_{ab}$
(double-dashed line in Fig. \ref{FD}(b)) of the boson propagator can
be found after evaluating the $6 \times 6$ boson self-energy
matrix $\Pi_{ab}$. $\Pi_{ab}$  may be evaluated by Feynman rules
through the diagrams in Fig. \ref{FD}(b).

In the present summary there is no mention of the ghost fields.
They were already treated in previous papers and the only role
they play is to cancel the infinities given by the two diagrams
shown in Fig. \ref{FD}(b).

From the $N$-extended
completeness condition it may be seen that the charge operator
$X^{00}$ is $O(N)$, while the operators $X^{pp}$ are $O(1)$.
This fact will have the physical consequence that PIH
weakens the effective spin interaction compared to that
one related to the
charge degree of freedom.

The component $D_{RR}$ (component (1,1)) of the $6 \times 6$ boson
propagator
is related to the charge-charge correlation function $\chi^c$ by

\beq \label{chi}\chi^c(q,\nu_n)=N\left( \frac{\delta}{2} \right)^2
D_{RR}(q,\nu_n). \eeq

\noindent In Ref. \onlinecite{Foussats}, \onlinecite{Foussats1} we have pointed
out that, in $O(1)$, the charge-charge correlation function shows
the presence of collective peaks above the particle-hole
continuum.

In what follows self-energies and one-particle
spectral functions are calculated by means of the Feynman rules.
The Green's function (\ref{G0}) corresponds to the $N$-infinite
propagator which includes no dynamical corrections; these appear
at higher order in the $1/N$ expansion. For obtaining spectral
densities,  the self-energy $\Sigma$ is calculated. Using the
Feynman rules, the total self-energy in $O(1/N)$ is obtained
adding the contribution of the two diagrams shown in Fig.
\ref{FD}(c). The analytical expression for $\Sigma$, for a given
channel $p$, results:

\begin{equation}
\Sigma({\mathbf{k}},i\omega_{n})=
\Sigma^{(1)}({\mathbf{k}},i\omega_{n})+
\Sigma^{(2)}({\mathbf{k}},i\omega_{n}) \; ,
\end{equation}

\noindent where

\begin{eqnarray}\label{Sigma1}
\Sigma^{(1)}({\mathbf{k}},i\omega_{n})&=&\frac{1}{N_{s}}\sum_{{\mathbf{q}}
,\nu_{n}}\Lambda^{pp}_{a}
\;D_{ab}({\mathbf{q}},i\nu_{n}) \nonumber\\
&\times& \; G_{pp}^{(0)} ({\mathbf{k-q}},i(\omega_{n}-\nu_{n})) \;
\Lambda^{p p}_{b}
\end{eqnarray}

\noindent and 

\begin{eqnarray}\label{Sigma2}
\Sigma^{(2)}({\mathbf{k}},i\omega_{n})&=&\frac{1}{N_{s}}\sum_{{\mathbf{q}}
,\nu_{n}}\Lambda^{pp}_{ab}\;
D_{ab}({\mathbf{q}},i\nu_{n}).
\end{eqnarray}

\noindent The sum over repeated indices $a$ and $b$ is assumed.
The renormalized boson propagator $D_{ab}$ plays a similar role as
the phonon propagator when dealing with the electron-phonon
interaction in simple metals. Therefore, in the calculation of
$\Sigma({{\bf{k}}},\omega)$ through $O(1/N)$, enter  the band
structure effects and collective effects associated with the
charge degrees of freedom (see Eq. (\ref{chi})).
Using the spectral representation for the boson field, $D_{ab}$,
we obtain

\begin{eqnarray}\label{Sigma11}
\Sigma^{(1)}({\mathbf{k}},i\omega_{n})&=&\frac{1}{2\pi N_{s}}\int
d\nu \sum_{{\mathbf{q}}
,\nu_{n}}\Lambda_{a}^{pp}
\frac{B^{ab}({\mathbf{q}},\nu)}{i\nu_{n}-\nu}\;
\Lambda_{b}^{pp} \nonumber \\
&\times& G_{pp}^{(0)}({\mathbf{k-q}},i(\omega_{n}-\nu_{n})),
\end{eqnarray}

\begin{equation}\label{Sigma21}
\Sigma^{(2)}({\mathbf{k}},i\omega_{n})=\frac{1}{2\pi N_{s}}\int
d\nu \sum_{{\mathbf{q}}
,\nu_{n}} \Lambda_{ab}^{pp} \;\frac{B^{ab}
({\mathbf{q}},\nu)}{i\nu_{n}-\nu} ,
\end{equation}

\noindent where

\begin{equation}\label{B}
B^{ab}({\mathbf{q}},\nu)=-2\lim_{\eta\rightarrow0}\; {\mathrm{Im}}[D_{ab}
({\mathbf{q}},i\nu_{n}\rightarrow\nu+i\eta)].
\end{equation}

After performing the Matsubara sum and the
analytical continuation $i\omega_n=\omega+i\eta$,
the imaginary part of $\Sigma$ is

\begin{eqnarray}\label{SigmaIm}
    {\mathrm{Im}}\Sigma({\mathbf{k}},\omega) &=& \frac{1}{2 N_{s}}
\sum_{{\mathbf{q}}} h_{a}(k,q,\omega-E_{{k-q}}) \nonumber\\
&& B^{ab}({\mathbf{q}},\omega-E_{{k-q}}) \; h_{b}(k,q,\omega-E_{{k-q}}) \nonumber\\
&&\times [n_{F}( -E_{k-q}) +n_{B}(\omega-E_{{k-q}})]
\end{eqnarray}

\noindent where $n_B$ is the Bose factor,
and the 6-component vector $h_{a} (k,q,\nu)$ is

\begin{widetext}
\begin{eqnarray*}
h_{a} (k,q,\nu)
&=&\left( \frac{}{} \right. \frac{2E_{k-q}+\nu+2\mu}{2}
  + 2\Delta \sum_{\eta} \cos(k_\eta-\frac{q_\eta}{2}) \; ;
1 \; ; -2 \; \Delta \; \cos(k_{x}-\frac{q_{x}}{2}) \; ;
-2\;\Delta\;\cos(k_{y}-\frac{q_{y}}{2}) \; ; \; \nonumber \\
&& 2 \; \Delta \; sin(k_{x}-\frac{q_{x}}{2}) \; ; \;
2 \; \Delta \; sin(k_{y}-\frac{q_{y}}{2}) \left. \frac{}{} \right).
\end{eqnarray*}
\end{widetext}

It is interesting to show the more compact result  for the case
$J=0$:

\begin{eqnarray}\label{SigmaIm0}
    {\mathrm{Im}}\Sigma({\mathbf{k}},\omega)&=& \frac{1}{2 N_{s}}
\sum_{{\mathbf{q}}} \left\{ \Omega^{2} \;
 B^{RR}({\mathbf{q}},\omega-E_{{k-q}}) \right. \nonumber\\
&& \hspace{-1cm} + \; 2\;\Omega \;
B^{\lambda R}({\mathbf{q}},\omega-E_{{k-q}})
+ \left.
B^{\lambda \lambda}({\mathbf{q}},\omega-E_{{k-q}}) \right\} \nonumber\\
&& \hspace{-1cm}\times \left(n_{F}(-E_{k-q}) +
n_{B}(\omega-E_{{k-q}})\right) \; ,
\end{eqnarray}

\vspace{0.5cm}

\noindent where $\Omega=(E_{k-q}+2\mu+\omega)/2$.
Using the Kramers-Kronig relations  can be determined $\mathrm{Re}
\Sigma({{\bf{k}}},\omega)$ from $\mathrm{Im}
\Sigma({{\bf{k}}},\omega)$ and compute the spectral function
$A({\bf k},\omega)=-\frac{1}{\pi}
 {\rm Im}G({\bf k},\omega)$ as

\begin{eqnarray}\label{A}
A({\bf k},\omega)= -\frac{1}{\pi}\frac{{\rm Im}\Sigma({\bf
k},\omega)}{(\omega- E_{{\bf k}}-{\rm Re}\Sigma({\bf k},\omega))^2
+ {\rm Im}\Sigma({\bf k},\omega)^2} \label{ak}
\end{eqnarray}

The self-energy is calculated using the propagator $G({\bf
k},\omega)$ for the $f$-operators which are proportional to the
fermionic Hubbard operators and then, they can not be related to 
usual fermions.

\section{Comparison with slave-boson}

While many papers on SB have been published on the mean field
level there are few calculations including fluctuations above the
mean field which are necessary for the estimation of spectral
functions. This shows that, in spite of the popularity of the SB
method, it is not trivial to implement this kind of calculations.
Even if PIH results have, at the mean field level,  a close connection 
with SB (see
Refs. \onlinecite{Foussats}, \onlinecite{Foussats1}) it is relevant to compare
both approaches beyond the mean field level. 
To our knowledge, 
there is only
one paper where spectral functions 
have been 
calculated in the framework of SB \cite{KotliarS}. 
In that paper,
their authors present results for $J=0.3$ and doping $\delta=0.20$
for three different ${\bf{k}}$-points on the BZ.

\begin{figure}
\begin{center}
\setlength{\unitlength}{1cm}
\includegraphics[width=8cm,angle=0]{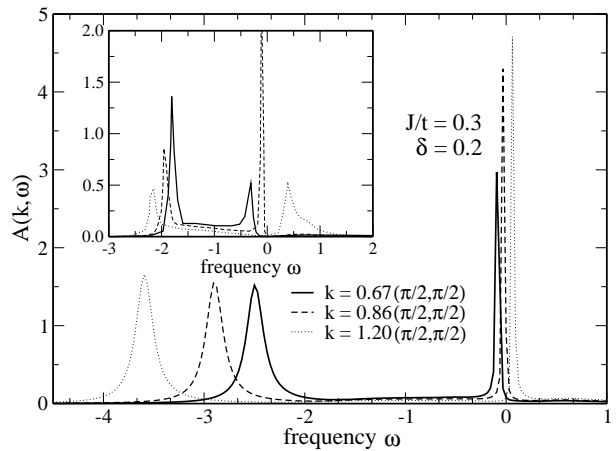}
\end{center}
\caption{ Solid, dashed, and doted lines are the spectral
functions for ${\bf{k}}=0.67(\pi/2,\pi/2)$,
${\bf{k}}=0.86(\pi/2,\pi/2)$ and ${\bf{k}}=1.2(\pi/2,\pi/2)$
respectively. 
The frequency $\omega$ is in units of $t$ which is considered to be 1.
For comparison with SB, in the inset we have
included the results from Ref. \onlinecite{KotliarS}.}
\label{Avsw}
\end{figure}

In Fig. \ref{Avsw} we present PIH spectral functions
$A({\bf{k}},\omega)$. The calculation was done for $J=0.3$,
$\delta=0.20$ and for ${\bf{k}}=0.67(\pi/2,\pi/2)$,
${\bf{k}}=0.86(\pi/2,\pi/2)$ and ${\bf{k}}=1.2(\pi/2,\pi/2)$,
which are exactly the same conditions of Fig. 3 in Ref.
\onlinecite{KotliarS}. In the inset of Fig. \ref{Avsw} we have
included SB results for comparison.
As can be seen, the spectral functions have some similarities with
SB. We have obtained a low energy peak and a pronounced structure
at large binding energy. The low energy peak is the quasiparticle
(QP). The other features, at large binding energy, are incoherent
spectra (IS). For ${\bf{k}} \sim 0.86(\pi/2,\pi/2)$ the QP peak
crosses $\omega=0$, where the chemical potential is located.
In spite of similarities between present results and those of Ref.
\onlinecite{KotliarS} there are some differences: a) Our IS is
located at  binding energy larger than in SB. For instance, for
${\bf{k}}=0.86(\pi/2,\pi/2)$, IS is at $\omega\sim-3t$ while the
corresponding one in SB is at $\omega\sim -2t$. b) Our QP peak is
less dispersing than in SB. As it is well known, self-energy
effects depresses Fermi velocities ($v_F^*$)  respect to the bare
one ($v_F^{bare}$), $v_F^*=Z v_F^{bare}$, where $Z=(1-\frac{\partial
\Sigma}{\partial \omega})$ is the QP weight therefore, it is
concluded that our self-energy effects are larger than in SB.

In SB there are three different
$\Sigma$'s \cite{KotliarS}: $\Sigma_n$, $\Sigma_a$ and
$\Sigma_{inc}$. In $\Sigma_n$,  bosons are condensed, in
$\Sigma_a$ one boson is condensed and the other fluctuates, and in
$\Sigma_{inc}$ both bosons fluctuate. These complications are due
to the decoupling scheme used in SB, so beyond mean field level it
is necessary to convolute spinons and bosons for reconstructing
the original $X$-operators. As in the PIH approach there is no any
{\it{a priori}} decoupling and we work directly with the Hubbard
operators,  we  have only one self-energy $\Sigma$ given by Eq.
(\ref{Sigma1}). A detailed description of the present
self-energies for different $\delta$ and $J$ is given in section
V.

After comparing spectral functions with some available
SB results and, in spite of some similarities, we found
important differences in the self-energy effects between both methods.
The existence of only a few SB
results beyond mean field level may be closely related to the
decoupling scheme which leads to a gauge field theory, making hard
the implementation of the approach. We hope that PIH be useful and
a complement of the SB calculations.

\section{Comparison with exact diagonalization}

As pointed out in Sec. II, PIH approach weakens spin fluctuations
over charge fluctuations. For instance, at leading order, while
there are collective effects in the charge channel, the spin
channel exhibits the characteristic form of a Pauli paramagnet
where the electronic band is renormalized by correlations
\cite{Lew,Foussats1}. So that, at $O(1/N)$, the
self-energy does not contain collective effects, like magnons in
the spin channel. Intuitively one may think that the method will
be better for large doping than for low doping. However,
 the exact role played by charge and
magnetic excitations as a function of doping, in the $t-J$ model,
is one of the key points for understanding the physics of
cuprates. Many mechanisms have been proposed. Some of them
privilege charge while others privilege magnetism.

In order to test the reliability of our results as a function of doping, in this
section, we compare {\it qualitatively} present spectral functions with
calculations obtained using Lanczos diagonalization. For this
purpose we have performed Lanczos diagonalization \cite{Riera} on
the $4\times 4$ lattice for $\delta=0.75,\; 0.5$ and $0.3125$,
with $J=0.3$.

As an example, we will explicitly compare some ${\bf {k}}$-points
for several dopings 
leaving to the reader the analysis of the
other ${\bf {k}}$'s.\\

{\bf a) Results for doping $\delta=0.75$}\\

\begin{figure}
\begin{center}
\setlength{\unitlength}{1cm}
\includegraphics[width=8.5cm,angle=0]{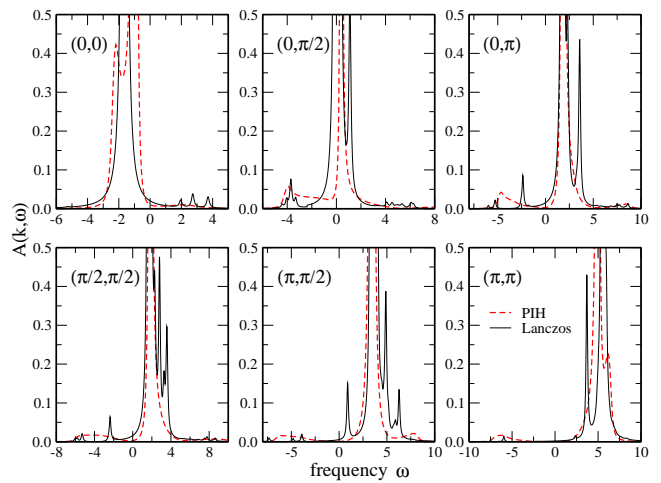}
\end{center}
\caption{
(Color online) Comparison between spectral function
calculations obtained by Lanczos
diagonalization for all the allowed ${\bf{k}}$-points of the $4\times 4$ lattice and
results obtained with the PIH method.
The parameters are $J=0.3$ and $\delta=0.75$.
The ${\bf{k}}$-points are presented between parenthesis in each frame. 
The frequency $\omega$ is in units of $t$.
In order to plot PIH and Lanczos results on the same figure
we have used an arbitrary scale for the $y$-axis.
}
\label{FE75}
\end{figure}

This doping corresponds to $12$ holes in the $4\times 4$ lattice.
There is good agreement between Lanczos and PIH (Fig. \ref{FE75})
for the six ${\bf{k}}$-points allowed in the $4\times4$ lattice.
For instance, for ${\bf{k}}=(0,\pi)$ both calculations show a QP
peak at around $\omega \sim 2t$ and IS at around $\omega \sim
-4t$. Lanczos diagonalization shows a small peak at $\omega \sim
3t$ which is not seen in PIH. However, PIH shows an asymmetric
shape of the QP peak which can be interpreted as an indication of the
additional structure observed in Lanczos. For ${\bf{k}}=(0,\pi/2)$
both, Lanczos and PIH show a QP peak near the Fermi level, and IS
at $\omega \sim -4t$. The additional peak that appears in Lanczos
at $\omega \sim 1.5t$ can be associated with the non-symmetrical
shape of the QP peak seen in PIH.

\vspace{0.5cm}

{\bf b) Results for doping $\delta=0.50$}\\

\begin{figure}
\begin{center}
\setlength{\unitlength}{1cm}
\includegraphics[width=8.5cm,angle=0]{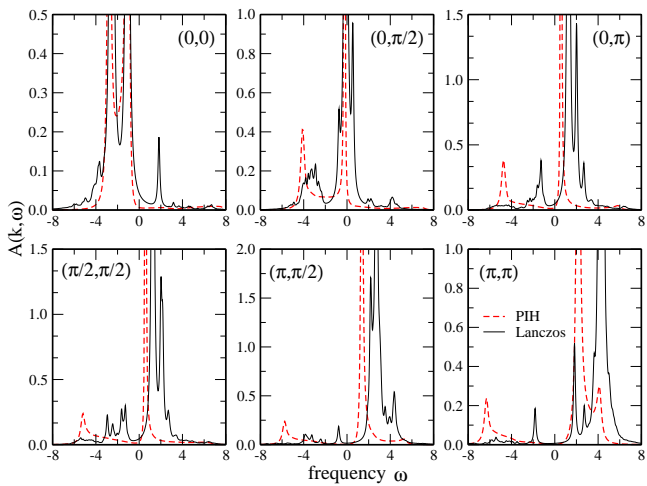}
\end{center}
\caption{ (Color online) Comparison between spectral function
calculations obtained by Lanczos diagonalization for all the
allowed ${\bf{k}}$-points of the $4\times 4$ lattice and results
obtained with the PIH method. The parameters are $J=0.3$ and
$\delta=0.5$. 
The ${\bf{k}}$-points are presented between parenthesis in each frame. 
The frequency $\omega$ is in units of $t$.
} \label{FE5}
\end{figure}

This doping corresponds to $8$ holes in the $4\times4$ lattice.
Results are presented in Fig. \ref{FE5} for Lanczos and PIH. With
decreasing doping from $0.75$ to $0.50$, both methods show more
IS. For instance, for ${\bf{k}}=(0,0)$ both calculations show two
well defined peaks below the Fermi level. The peak closer to the
chemical potential is the QP, and the peak near $\omega \sim -3t$
is of incoherent character. Both methods also present small IS for
$\omega > 0$.

For $\delta=0.5$, results  for $J=0$ were presented in Ref.
\onlinecite{Merino} in the context of organic materials where PIH
spectral functions were also compared with those obtained using
Lanczos diagonalization as a function
of the nearest neighbors Coulomb interaction $V$. \\

{\bf c) Results for doping $\delta=0.3125$}\\

\begin{figure}
\begin{center}
\setlength{\unitlength}{1cm}
\includegraphics[width=8.5cm,angle=0]{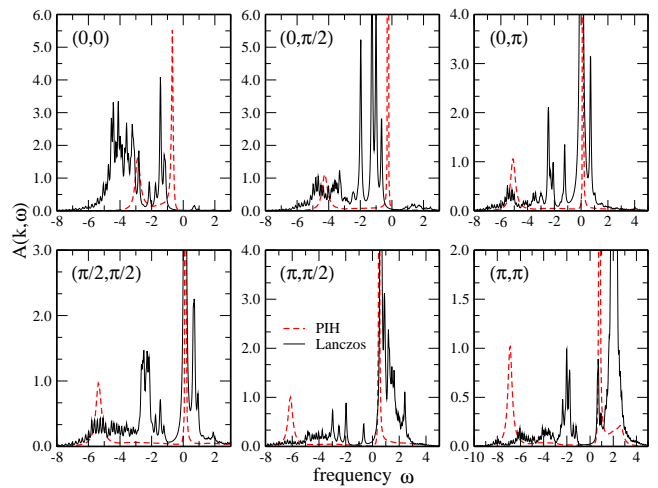}
\end{center}
\caption{
(Color online) Comparison between spectral function calculations obtained by Lanczos
diagonalization for all the allowed ${\bf{k}}$-points of the $4\times 4$ lattice and
results obtained with the PIH method.
The parameters are $J=0.3$ and $\delta=0.3125$.
The ${\bf{k}}$-points are presented between parenthesis in each frame. 
The frequency $\omega$ is in units of $t$.
}
\label{FE3125}
\end{figure}

This doping corresponds to $5$ holes in the $4\times4$ lattice.
The results are presented in Fig. \ref{FE3125} for Lanczos and PIH.
Lanczos and PIH present the QP peak near the Fermi level and
stronger IS than for $\delta=0.75$ and $\delta=0.5$. The
increasing of the IS is consistent with the fact that the QP
weight is $Z \sim 0.5$, lower than for the previous dopings where
for $\delta=0.75$ and $\delta=0.5$ is $Z \sim 0.9$ and $Z \sim
0.7$ respectively. The QP weight will be discussed in more detail
in next section.

For ${\bf{k}}=(0,0)$ both calculations show a QP peak below Fermi
energy. The peak that appears in PIH at $\omega \sim -3t$ is
broader in Lanczos and centered at $\omega \sim -4t$. For
${\bf{k}}=(\pi,\pi/2)$ both calculations show a QP peak above
Fermi level and IS on the top of this peak. The well pronounced
structure obtained in PIH at $\omega \sim -6t$ seems to be missing
in Lanczos. Instead, it shows an  homogeneously distributed IS
below the Fermi level up to large binding energies of the order of
$-7t$. In the frequency range $-5t <\omega< -1t$ both methods show
IS. For ${\bf{k}}=(0,\pi)$ both methods present a QP near the
Fermi level and IS at $\omega \sim -5t$. Between those two
features it is possible to see IS (in the form of several peaks in
Lanczos)
in both calculations. \\

For the three studied dopings both methods show that while the QP
peak disperses through the Fermi surface (FS), the edge of the IS
moves in opposite direction. This result was obtained previously by
Stephan and Horsch\cite{Stephan}. In Ref. \onlinecite{Stephan} the
authors also studied spectral functions for the $t-J$ model at
moderate doping $\delta\sim 0.1$ by means of exact
diagonalization. That paper  presents strong evidences for a large
FS for moderated doping levels. This result gives an additional
support for our bare band
$E_k=-2(t\;\delta/2+\Delta)(cos\,k_x+cos\;k_y)-\mu$.

We conclude that PIH and Lanczos results, for the three studied
$\delta$ values (which cover a broad range of doping),  fairly
agree considering the different nature of both methods. The above
comparison gives some confidence to our self-energies. The
self-energy has additional information such as relaxation times
$1/\tau$, quasiparticle weight $Z$ (effective mass increasing)  
which can not be directly obtained from Lanczos
diagonalization.

Decreasing doping , Lanczos
and PIH both show  band narrowing. The narrowing is stronger in
PIH than in Lanczos. For instance, for $\bf{k}=(\pi,\pi)$ (Fig.
\ref{FE5}), while the Lanczos QP peak is at $\omega \sim 4t$, the
PIH QP peak is at $\omega \sim 2t$. For $\bf{k}=(\pi,\pi)$ (Fig.
\ref{FE3125}),  while Lanczos QP peak is at $\omega\sim 2t$ it is
at $\omega \sim 1t$ in PIH. For $\delta=0.75$ (Fig. \ref{FE75})
there is good agreement in the QP and IS energy positions for each
$\bf{k}$. This discussion is important in the light of
$ARPES$ experiments in cuprates\cite{zhou, fink} which show Fermi
velocities rather independent of doping (see also Ref.
\onlinecite{Sorella} for discussions). In  PIH  the strong
narrowing is mainly due to the factor $\delta/2$ in the electronic
dispersion which strongly weakens the $t$-term.

\section{
Self-energy renormalizations
}

In this
section, we present a detailed self energy and spectral functions
calculations from PIH.

\begin{figure}
\begin{center}
\setlength{\unitlength}{1cm}
\includegraphics[width=7cm,angle=0]{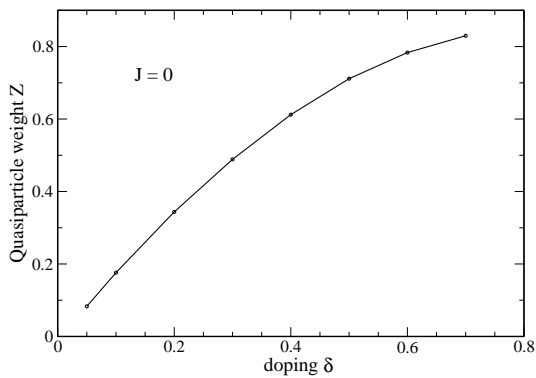}
\end{center}
\caption{
Quasiparticle weight $Z$ as a function of doping for $J=0$.
}
\label{Z0}
\end{figure}

Fig. \ref{Z0} shows the QP weight $Z$ as a function of doping for
$J=0$. As it was shown in Ref. \onlinecite{Foussats1}, for $J=0$
the homogeneous Fermi liquid (HFL) remains stable for all
$\delta$. For each doping, we have found that the self-energy is
very isotropic on the FS (see below), making the QP weight rather
constant.
In  Fig. \ref{Z0} $Z \rightarrow 0$ when $\delta \rightarrow 0$.
For small $\delta$, $Z \sim 1.4 \; \delta$, which is very close to
the observed $ARPES$ behavior in $LSCO$\cite{Yoshida}. As $Z$
remains finite for $\delta >0$, present calculation predicts a
Fermi liquid (FL) behaviour. Fig. \ref{Z0} also shows that $Z
\rightarrow 1$ when $\delta \rightarrow 1$ as expected for an
uncorrelated system.

\begin{figure}
\begin{center}
\setlength{\unitlength}{1cm}
\includegraphics[width=7cm,angle=0]{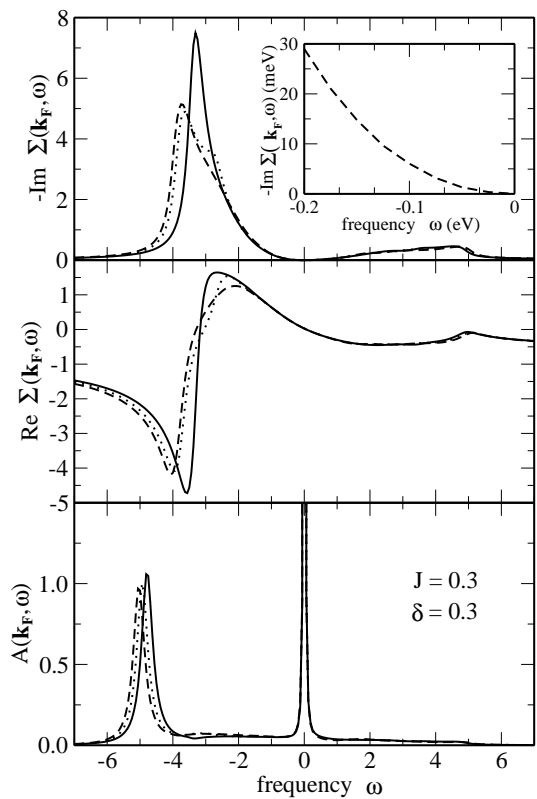}
\end{center}
\caption{
$-Im \Sigma({\bf{k_F}},\omega)$ (upper panel), $Re
\Sigma({\bf{k_F}},\omega)$
(middle panel) and spectral functions (lower panel) for $J=0.3$,
$\delta=0.30$ and three different points on the FS.
Solid line, dashed line and dotted line are results
for the Fermi point in the $(1,1)$, $(1,0)$ directions and between these two
respectively.
The frequency $\omega$, $Re \Sigma$ and $Im \Sigma$ are in units of $t$.
Inset: $-Im \Sigma({\bf{k_F}},\omega)$
for ${\bf{k_F}}$ in the $(11)$-direction in units of $t=0.4eV$.
}
\label{IRA3125}
\end{figure}

\vspace{0.5cm}

Let us discuss the case $J=0.3$. Fig. \ref{IRA3125} shows
$\Sigma$, for $\delta=0.3$, for three well separated
${\bf{k}}$-vectors on FS. One of the ${\bf{k}}$ is chosen in the
$(11)$-direction of the BZ, other in the $(10)$-direction and the
third in between both. The upper and the middle panel of
Fig. \ref{IRA3125} show the ${\mathrm{Im}} \Sigma$ and
${\mathrm{Re}} \Sigma$ respectively.

As shown in Fig. \ref{IRA3125}, PIH predicts a rather isotropic
self-energy on the FS. On the other hand, for each ${\bf{k}}$,
$\Sigma({\bf{k}},\omega)$ is very asymmetric respect to $\omega=0$
which can be interpreted as a consequence of the difference
between addition and removal of a single electron in a correlated
system. Near $\omega=0$, ${\mathrm{Im}} \Sigma({\bf{k}}_F,\omega)
\sim \omega^2$, showing FL behavior. On the other hand,
${\mathrm{Re}} \Sigma({\bf{k}}_F,\omega)$ shows, at $\omega=0$, a
negative slope which is also a characteristic of a Fermi liquid.

Inset of Fig. \ref{IRA3125} shows a plot of $-{\mathrm{Im}}
\Sigma({\bf{k_F}}, \omega)$ for $\omega < 0$ for ${\bf{k}}_F$ in
the $(11)$-direction. We have used $t=0.4\;eV$ \cite{Reviwes}. In
the range $-200 \;meV < \omega< 0$, $-{\mathrm{Im}}
\Sigma({\bf{k}},\omega)$ does not saturate as in Fig. 1 of Ref.
\onlinecite{zhou}. The no saturation of $-{\mathrm{Im}}
\Sigma({\bf{k}},\omega)$, up to an energy  scale of the order of
$-200\;(-300) \;meV$, is well established in cuprates and clearly
it can not be explained by phonons. In addition to this feature,
Fig. 1 of Ref. \onlinecite{zhou} shows the presence of an
additional energy scale of the order of $60-70 \;meV$ which is
associated with the kink observed in $ARPES$  \cite{Reviwes}. This
small energy scale is not seen in our $-{\mathrm{Im}}
\Sigma({\bf{k}},\omega)$. Whether the kink is due to magnetic
excitations or additional degrees of freedom like phonons, is still 
controversial  \cite{Reviwes,Zeyher3}. For instance, Yunoki {\it
et~ al.} \cite{Sorella}, using variational MC, found no evidence
for the kink in the context of the pure $t-J$ model and, on the
other hand, FLEX calculations for the Hubbard model suggest that
the interaction between QP and spin fluctuations leads to the kink
(see Ref. \onlinecite{Manske} and references therein). As
mentioned in previous sections the PIH approach weakens spin
fluctuations over charge fluctuations, which means that $1/N^2$
self-energy corrections should be calculated in order to study the
kink,  if originated by magnetic excitations.

Finally, as expected from the results shown in the upper and
middle panels, the lower panel of Fig. \ref{IRA3125} shows that,
for the three mentioned ${\bf{k}}$-vectors, the corresponding
spectral functions are isotropic.

As it was shown in Ref. \onlinecite{Foussats1}, in agreement with
previous calculations  \cite{Marston,Lubenski,Cappelluti}, for
$J=0.3$, the system presents a flux phase (FP) (also called
$d$-density wave (DDW)\cite{Chakra}) for $\delta < \delta_c \sim
0.14$. FP was interpreted as a candidate for the pseudogap phase
of cuprates  \cite{Cappelluti,Chakra}. Thus, it is important
to study self-energy corrections approaching  the FP instability
from the HFL phase ($\delta \rightarrow \delta_c$ from above).
Similar calculations to those for $\delta=0.3$ show, for $\delta
\gtrsim \delta_c \sim 0.14$, isotropic self-energy effects on the
FS.

\begin{figure}
\begin{center}
\setlength{\unitlength}{1cm}
\includegraphics[width=7cm,angle=0]{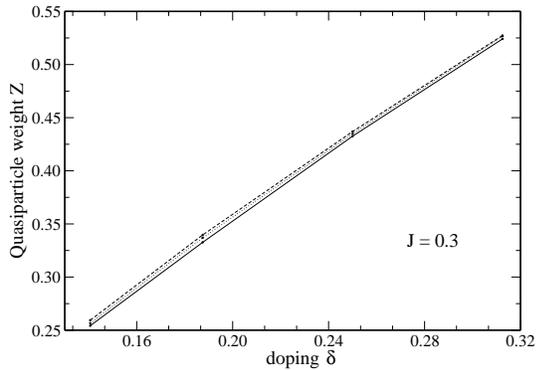}
\end{center}
\caption{ Quasiparticle weight $Z$ versus doping for the three
${\bf{k}}$-points on the FS discussed in Fig. \ref{IRA3125}. As
discussed in the text there is no indication of the proximity to
the DDW phase. The calculations run from $\delta>\delta_c \sim
0.14$ where the HFL is stable. } \label{Z3}
\end{figure}

Fig. \ref{Z3} shows $Z$ versus $\delta$ for three
${\bf{k}}$-points chosen as in Fig. \ref{IRA3125}; each one of
them on its corresponding FS for each doping. Results are for
$\delta <0.3$. The QP weight results very isotropic on the FS even
for doping near $\delta_c$. According to our results the
anisotropy, between $X$-point ($(10)$-direction) and nodal point
($(11)$-direction), observed in $ARPES$ spectra in cuprates, can not
be interpreted as originated by self-energy effects. This is
close to the recent interpretation by Kaminski {\it et~ al.}
\cite{Kami} where the scattering rate was found to be composed by
an isotropic inelastic term and a highly anisotropic elastic term
which correlates with the anisotropy of the pseudogap. In our
case, the ${\rm{Im}} \Sigma$  can be interpreted as the inelastic
contribution to the scattering rate and the opening of the flux
phase, which is mainly of static character
\cite{Cappelluti,Foussats1}, as the elastic term. (A close
comparison with $ARPES$ experiments in cuprates, which needs a
better FS as given by the $tt'-J$ model,  is in progress).

It is important to discuss the reason for the non strong influence
of the flux instability on the self-energy. As discussed in Refs.
\onlinecite{Foussats1}, \onlinecite{Cappelluti}, flux phase is mainly of static
and $d$-wave symmetry character and it is weakly coupled to the
charge sector. Being our self-energy dominated by charge
fluctuations (see below), $\Sigma$ does not strongly prove the
proximity to the DDW. In terms of Ref. \onlinecite{Chakra}, 
the proximity to the DDW is hidden for the
one-particle spectral densities. This is in contrast with the
self-energy behaviour in the proximity of the usual charge density
wave (CDW) instability. In Ref. \onlinecite{Merino} it was shown
that the QP weight $Z$ is strongly affected when the system
approaches the CDW phase.

A discussion about the excitations that, interacting with
electrons, cause the self-energy renormalizations is necessary. In
the usual many body language, the self-energy can be expressed in
terms of the relevant quantity $\alpha^2 F(\omega)$\cite{Mahan},
where the notation is chosen in  a way such that $F(\omega)$ gives
information on the density of states of a boson interacting with
the electrons, and $\alpha^2$ about the coupling. In usual metals,
$\alpha^2 F(\omega)$ contains information of the electron-phonon
interaction averaged over the FS. For simplicity, we will discuss the  
$J=0$ case. Eq. (\ref{SigmaIm0}) is conveniently
written for reading $\alpha^2 F(\omega)$. In the first term of the
second hand side of Eq. (\ref{SigmaIm0}) we can interpret
$B^{RR}=-2 {\rm{Im}} D_{RR}$ as the spectral function of the boson
mediating the interaction, while the remaining squared factor,
$\Omega^2$, as the coupling. As discussed in Sec. II, $D_{RR}$
corresponds to the charge-charge correlation function (Eq. (9)).

\begin{figure}
\begin{center}
\setlength{\unitlength}{1cm}
\includegraphics[width=7cm,angle=0]{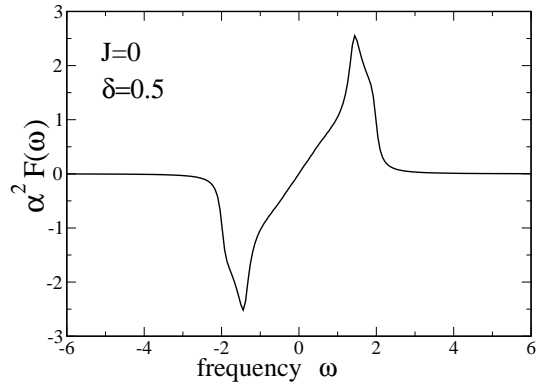}
\end{center}
\caption{
$\alpha^2F(\omega)$ as extracted from Eq. (\ref{SigmaIm0}) for
$\Sigma$.
As discussed in the text, $\alpha^2 F(\omega)$ is mainly dominated by
collective excitations of the charge character. The frequency $\omega$ and $\alpha^2F$ 
are in units of $t$.
}
\label{a2F}
\end{figure}

Fig. \ref{a2F} shows  $\alpha^2 F(\omega)$ obtained from the first
term of Eq. (17). Following the discussion above, $\alpha^2
F(\omega)$ is proportional to the average on the FS of the charge
densities. Clearly, charge densities survives up to high energy
causing the large self-energy effects at large $\omega$. Since
charge densities, in $O(1)$, present collective peaks at the top
of the particle-hole continuum\cite{Foussats1} both,  the
collective excitations and the continuum, contribute to $\alpha^2
F(\omega)$. For instance, the pronounced structure at $\omega \sim
\pm 1.7$ in Fig. \ref{a2F} is mainly due to collective
fluctuations. The interpretation of the last two terms of the
second hand side of Eq.(\ref{SigmaIm0}) is less direct and they are
proper of our strong coupling perturbative approach. However, they
are also dominated by collective excitation of charge character
arising from
the inversion of the matrix $D$. \\

{\it Sum rule:} Before closing we will discuss the spectral
function sum rule $\int d\omega \;A(\bf{k},\omega)$. In the
framework of the  $t-J$ model the sum rule is $<X^{00}>+<X^{\sigma
\sigma}>=(1+\delta)/2$. Using the relation $X^{0\sigma}=\sqrt{N
r_0}\; f_p$, in the limit $N=2$, our sum rule is $\delta$
($<X^{00}>$), therefore, PIH misses a contribution $<X^{\sigma
\sigma}>=(1-\delta)/2$ making the situation better for large than
for low doping. It is important to discuss about a possible origin
of this discrepancy. As was pointed out in previous papers
\cite{Foussats4,Foussats5}, in order to guarantee the commutation
rules for $X$-operators not all the multiplication rules can be
satisfied. For instance, in Ref. \onlinecite{Foussats4} we have
studied the spin system using the four $X$-operators $X^{\sigma
\sigma'}$ and showed that the formulation leads to the well known
coherent state path integral representation for
spins\cite{Fradkin}. The fact that this representation is better
for large than for small $S$ (Ref. \onlinecite{Fradkin}) was
understood, in Ref. \onlinecite{Foussats4}, as a consequence that
not all the multiplication rules are fulfilled. It is worthy to
note that the formalism in Ref. \onlinecite{Foussats4} reproduces
the spinless fermion case when they are written using $X$-operator
representation. In the present case we deal with the $t-J$ model
and in order to satisfy the commutation rules, the formalism
requires the constraint $X^{\sigma0} X^{0\sigma'}=X^{00}X^{\sigma
\sigma'}$ \cite{Foussats,Foussats5} which reproduces the exact
multiplication rule $X^{\sigma0} X^{0\sigma'}=X^{\sigma \sigma'}$
in the limit $X^{00} \rightarrow 1$ ($\delta \rightarrow 1$)
making the representation better for large than for low doping. In
mathematical terms, our expansion seems to be appropriate for both
large $N$ and large $\delta$. This is closely related to the fact
that the formulation weakens spin over charge fluctuations. The
band narrowing, discussed in section IV, is possibly connected
with this discussion if a spin term $<X^{\sigma \sigma}>$ also
contributes to the bandwidth. The solution of this very hard
theoretical problem, and the knowledge of how important it is as a
function of doping on different physical quantities, requires not
only to make an effort on formal level, but also, at the same
time, confronting results with different methods.

\section{Discussions and Conclusions}

The recently developed path-integral large-$N$ approach for
Hubbard operators, PIH, was used for calculating self-energy
corrections and spectral functions, including fluctuations above
mean field solution  of the $t-J$ model.

Similarities and differences with SB were discussed in section III.
To gain confidence on our calculation, comparisons of spectral
functions with Lanczos results for $J=0.3$ and for doping
$\delta=0.75$, $\delta=0.5$ and $\delta=0.3125$ were performed in
section IV. We found fair agreement for each ${\bf{k}}$ on the BZ.
PIH self-energies and spectral functions for different
$J$ and $\delta$ have been investigated in Sec. V. The general
characteristics of the self-energy are:

a) Around $\omega=0$, ${\rm{Im}} \Sigma(\omega) \sim \omega^2$ which is
characteristic of a FL behavior. This is in agreement
with the 
negative slope of ${\rm{Re}} \Sigma(\omega)$ at $\omega=0$.

b) $\Sigma(\omega)$ is very asymmetric with respect to $\omega=0$,
indicating the difference between addition  and removal of one
electron in a strongly  correlated system.

c) $\Sigma(\omega)$ has large structures  at large negative $\omega$
of the order of a few $t$.

For $J=0$, we have shown that $Z$ decreases monotonically as 
$\delta \rightarrow 0$ 
remaining 
finite for $\delta >0$.
For small $\delta$, $Z \sim 1.4 \; \delta$. 
As expected, in the uncorrelated limit,
$Z\rightarrow 1$ for $\delta \rightarrow 1$. 

We have also studied spectral functions along the FS for different
$\delta$ and $J=0.3$. For this case, the HFL is unstable against a
DDW phase for doping $\delta < \delta_c \sim 0.14$. Since DDW
phase was interpreted as a candidate for describing the pseudogap
state in cuprates, we have studied the behavior of the self-energy
along the FS when approaching the DDW instability. It has been
found that self-energy effects and  spectral functions are very
isotropic along the FS even for doping close to $\delta_c$.

In Sec. V we have discussed the nature of the excitations,
which interacting with the charge carriers produce the self-energy
renormalizations. Charge excitations, dominated by collective
effects, are the main contribution to $\alpha^2 F(\omega)$. As
collective charge fluctuations live on a large energy scale, they
are the responsible for the large self-energy effects at large
energy, producing the reduction of the quasiparticle weight and
transferring spectral weight to the incoherent spectra at large
binding energy.

PIH method seems to be a suitable alternative for calculating
spectral functions in the $t-J$ model, moreover  it can be used
independently or as a complement to other calculations as well.

\noindent{\bf Acknowledgments}

We thank L. Manuel, J. Merino, A. Muramatsu, A. Trumper  and R. Zeyher
for stimulating discussions and H. Parent
for critical reading of the manuscript.



\begin{thebibliography}{99}

\bibitem{Muller} J.G. Bednorz and K.A. M\"uller,
Zeit Phys. B {\bf64}, 189 (1986).

\bibitem{Anderson} P.W. Anderson, {\it{The Theory of Superconductivity in
High-$T_c$ Cuprates}} (Princeton University Press, Princeton,
1997).

\bibitem{Reviwes} A. Damascelli, Z-X. Shen and Z. Hussain,
Rev. Mod. Phys. {\bf 75}, 473 (2003).

\bibitem{Horsch} G. Martinez and P. Horsch,
Phys. Rev. B {\bf 44}, 317 (1991).

\bibitem{Izyumov} A. Izyumov, Physics-Uspekhi {\bf 40}, 445 (1997).

\bibitem{Muramatsu} M. Brunner, F. Assaad and A. Muramatsu,
Phys. Rev. B {\bf 62}, 15480 (2000).

\bibitem{Dagotto} E. Dagotto, Rev. Mod. Phys. {\bf 66}, 763 (1994).

\bibitem{Prelov} J. Jaklic and P. Prelovsek, Adv. Phys.
{\bf 49}, 1 (2000).

\bibitem{Foussats} A. Foussats and  A. Greco, Phys. Rev. B {\bf 65},
195107 (2002).

\bibitem{Arrigoni} E. Arrigoni, C. Castellani, M. Grilli, R. Raimondi and G. Strinati,
Phys. Rep. {\bf 241}, 291 (1994).


\bibitem{Foussats1} A. Foussats and  A. Greco, Phys. Rev. B {\bf 70},
205123 (2004).

\bibitem{Wang} Z. Wang, Int. Journal of Modern Physics B {\bf 6}, 155 (1992).

\bibitem{Zeyher} R. Zeyher and M. L. Kuli\'c, Phys. Rev. B {\bf 53}, 2850 (1996).


\bibitem{Merino} J. Merino, A. Greco, R. H. McKenzie, and M. Calandra,
Phys. Rev. B {\bf 68}, 245121 (2003).

\bibitem{Merino1} A. Greco, J. Merino, A. Foussats and R. H. McKenzie,
Phys. Rev. B {\bf 71}, 144502 (2005).

\bibitem{KotliarS} Z. Wang, Y. Bang and G.  Kotliar, Phys. Rev. Lett.
{\bf 67}, 2733 (1991).

\bibitem{Lew}  L. Gehlhoff and R. Zeyher,
Phys. Rev. B {\bf 52}, 4635 (1995).

\bibitem{Riera}
We thank J. Riera for usefull discussions and for
his desinterested offer of his Lanczos programs.


\bibitem{Stephan} W. Stephan and P. Horsch, Phys. Rev. Lett. {\bf 66},
2258 (1991).

\bibitem{zhou} X. J. Zhou, T. Yoshida, A. Lanzara, P. Bogdanov, S. Kellar,
K. Sherr, W. Yang, F. Ronning, T. Sasagawa, T. Kakeshita, T. Noda, H. Eisaki, S. Uchida,
 C. Lin, F. Zhou, J. Xiong, W. Ti, Z. Zhao, A. Fujimori, Z. Hussain and Z.-X. Shen,
Nature {\bf 423}, 398 (2003).

\bibitem{fink} J. Fink, S. Borisenko, A. Kordyuk, A. Koitzsch, J. Geck,
 V. Zabolotnyy, M. Knupfer, B. B\"uchner and H. Berger, cond-mat/0512307.

\bibitem{Sorella} S. Yunoki, E.
Dagotto and S. Sorella, Phys. Rev. Lett. {\bf 94}, 037001 (2005)

\bibitem{Yoshida} T. Yoshida, X. Zhou, T. Sasagawa, W. Yang, P. Bogdanov, A. Lanzara,
Z. Hussain, T. Mizokawa, A. Fujimori, H. Eisaki, Z.-X. Shen, T.
Kakeshita and S. Uchida, Phys. Rev. Lett. {\bf 91}, 027001 (2003).

\bibitem{Zeyher3} R. Zeyher and A. Greco, Phys. Rev. B {\bf 64},
140510 (R) (2001).

\bibitem{Manske} D. Manske, {\it{The Theory of Unconventional
Superconductors}}, (Springer-Verlag Berlin-Heidelberg, Germany
2004).

\bibitem{Marston} I. Affleck and J.B. Marston,
Phys. Rev. B {\bf37}, 3774 (1988).

\bibitem{Lubenski} D.C. Morse and T.C. Lubensky,
Phys. Rev. B {\bf 42}, 7994 (1990).

\bibitem{Cappelluti} E. Cappelluti and R. Zeyher, Phys. Rev. B {\bf 59},
6475 (1999).

\bibitem{Chakra} S. Chakravarty, R. B. Laughlin,
D. K. Morr, and Ch. Nayak, Phys. Rev. B {\bf 63}, 94503 (2001).

\bibitem{Kami} A. Kaminski, H. Fretwell, M. Norman, M. Randeria, S. Rosenkranz,
U. Chatterjee, J. Campuzano, J. Mesot, T. Sato, T. Takahashi, M.
Takano, K. Kadowaki, Z. Li and H. Raffy, Phys. Rev. B {\bf 71},
014517 (2005).

\bibitem{Mahan} G. Mahan,
{\it{Many-Particle Physics}} (Plenum Press, New York, 1981).

\bibitem{Foussats4} A. Foussats, A. Greco and O. S.
Zandron, Ann. Phys. (N.Y.) {\bf 275}, 238 (1999); {\bf 279}, 263
(2000).

\bibitem{Foussats5} A. Foussats, A. Greco, C. Repetto, O. P. Zandron and O. S.
Zandron, Journal of Physics A {\bf 33}, 5849 (2000).

\bibitem{Fradkin} E. Fradkin, {\it{Field Theories of Condensed Matter Systems}}
(Addison-Wesley, Reading, Massachusetts, 1991).


\end{thebibliography}
\end{document}